\documentclass[aps, prd,twocolumn,groupedaddress,showpacs,preprintnumbers]{revtex4}
\usepackage{epsf} 
\begin{document}

\preprint{MCGILL-03-06}
%\preprint{hep-th/0304168}
\preprint{hep-th/0304168}

\title{Towards a Stringy Resolution of the Cosmological Singularity}


\author{Damien A. Easson}%
	\email[Email:]{easson@hep.physics.mcgill.ca}
%\thanks{}
\affiliation{Department of Physics, McGill University, Montr{\'e}al, Qu{\'e}bec, H3A 2T8, Canada}


\date{\today}

\begin{abstract}
We study cosmological solutions to the low-energy effective action of heterotic
string theory including possible leading order $\alpha'$ corrections and a potential
for the dilaton. We consider the possibility that including such stringy corrections can 
resolve the initial cosmological singularity. Since the exact form
of these corrections is not known the higher-derivative terms are constructed so that they vanish
when the metric is de Sitter spacetime. The constructed terms
are compatible with known restrictions from scattering amplitude and string worldsheet
beta-function calculations. Analytic and numerical techniques
are used to construct a singularity-free cosmological solution. At late times and 
low-curvatures the metric is asymptotically Minkowski and
the dilaton is frozen. In the high-curvature regime the universe enters a de Sitter phase. 
\end{abstract}

%\keywords{}

\maketitle

% User definitions  
%--------------------------------------------------------------------  
\def\Box{\nabla^2}  
%---------------------------  
\def\ie{{\em i.e.\/}}  
\def\eg{{\em e.g.\/}}  
\def\etc{{\em etc.\/}}  
\def\etal{{\em et al.\/}}  
%----------------------------  
\def\S{{\mathcal S}}  
\def\I{{\mathcal I}}  
\def\mL{{\mathcal L}}  
\def\H{{\mathcal H}}  
\def\M{{\mathcal M}}  
\def\N{{\mathcal N}} 
\def\O{{\mathcal O}} 
\def\cP{{\mathcal P}} 
\def\R{{\mathcal R}}  
\def\K{{\mathcal K}}  
\def\W{{\mathcal W}} 
\def\mM{{\mathbf M}} 
\def\mP{{\mathbf P}} 
\def\mT{{\mathbf T}} 
\def\mR{{\mathbf R}}
\def\mS{{\mathbf S}}
\def\mX{{\mathbf X}}
\def\mZ{{\mathbf Z}}
%-----------------------------  
\def\eff{{\mathrm{eff}}}  
\def\Newton{{\mathrm{Newton}}}  
\def\bulk{{\mathrm{bulk}}}  
\def\brane{{\mathrm{brane}}}  
\def\matter{{\mathrm{matter}}}  
\def\tr{{\mathrm{tr}}}  
\def\normal{{\mathrm{normal}}}  
\def\implies{\Rightarrow}  
\def\half{{1\over2}}  
%------------------------------
\newcommand{\da}{\dot{a}}
\newcommand{\db}{\dot{b}}
\newcommand{\dn}{\dot{n}}
\newcommand{\dda}{\ddot{a}}
\newcommand{\ddb}{\ddot{b}}
\newcommand{\ddn}{\ddot{n}}
%------------------------------
\def\be{\begin{equation}}
\def\ee{\end{equation}}
\def\bea{\begin{eqnarray}}
\def\eea{\end{eqnarray}}
\def\bs{\begin{subequations}}
\def\es{\end{subequations}}
\def\g{\gamma}
\def\G{\Gamma}
\def\vp{\varphi}
\def\mpl{M_{\rm P}}
\def\ms{M_{\rm s}}
\def\ls{\ell_{\rm s}}
\def\lp{\ell_{\rm pl}}
\def\l{\lambda}
\def\gs{g_{\rm s}}
\def\d{\partial}
\def\co{{\cal O}}
\def\sp{\;\;\;,\;\;\;}
\def\spa{\;\;\;}
\def\r{\rho}
\def\dr{\dot r}
\def\dt{\dot\varphi}
\def\e{\epsilon}
\def\k{\kappa}
\def\m{\mu}
\def\n{\nu}
\def\om{\omega}
\def\tn{\tilde \nu}
\def\p{\phi}
\def\vp{\varphi}
\def\P{\Phi}
\def\r{\rho}
\def\s{\sigma}
\def\t{\tau}
\def\x{\chi}
\def\z{\zeta}
\def\a{\alpha}
\def\b{\beta}
\def\de{\delta}
\def\bra#1{\left\langle #1\right|}
\def\ket#1{\left| #1\right\rangle}
\newcommand{\stt}{\small\tt}
\renewcommand{\theequation}{\arabic{section}.\arabic{equation}}
\newcommand{\eq}[1]{equation~(\ref{#1})}
\newcommand{\eqs}[2]{equations~(\ref{#1}) and~(\ref{#2})}
\newcommand{\eqto}[2]{equations~(\ref{#1}) to~(\ref{#2})}
\newcommand{\fig}[1]{Fig.~(\ref{#1})}
\newcommand{\figs}[2]{Figs.~(\ref{#1}) and~(\ref{#2})}
\newcommand{\GeV}{\mbox{GeV}}

%-----------------------------------------------------------------------   
\def\ricci{R_{\m\n} R^{\m\n}}
\def\riemann{R_{\m\n\l\s} R^{\m\n\l\s}}
\def\triemann{\tilde R_{\m\n\l\s} \tilde R^{\m\n\l\s}}
\def\tricci{\tilde R_{\m\n} \tilde R^{\m\n}}
\def\La2{R_{\m\n\l\s} R^{\m\n\l\s} - \frac{1}{6} R^2}
\def\L2{\mathcal{L}_2}
\section{Introduction}  
%------------------------------------------------------------------------------  
\label{intro}
%------------------------------------------------------------------------------
Arguably, the most perplexing problem of modern cosmology is the initial singularity
problem. The theorems of Hawking and Penrose state that many of
the manifolds of General Relativity are geodesically incomplete~\cite{Hawking:1969sw}.
In particular, many solutions to the Friedmann-Robertson-Walker (FRW) isotropic and
homogeneous model of the universe contain singularities.
In order to gain a complete description of the early universe
a theory of quantum gravity is required.
In general, it is believed that such a theory may somehow resolve the 
initial singularity allowing us to obtain well-defined, finite solutions to calculations
of physical quantities.

Superstring theory is currently the best candidate for a theory of quantum gravity. 
It is therefore only natural to 
try to use string theory to probe the very early universe. Many of us are hopeful that
string theory (or M-theory) will lead to a consistent cosmological model capable of 
resolving the initial singularity~\footnote{A cosmological model based on string theory
that attempts to solve the initial singularity problem is the ``Brane Gas" model of 
\cite{Brandenberger:1988aj, Sakellariadou:1995vk, Alexander:2000xv, Brandenberger:2001kj, Easson:sj}. Some recent progress in 
string cosmology is reviewed in~\cite{Lidsey:1999mc}-\cite{Gasperini:2002bn}.}.  We are inspired, in part,
by the ability of stringy physics to resolve certain
singularities in time-independent orbifolds and 
conifolds~\cite{Liu:2002yd}-\cite{Greene:1995hu}. A full understanding of the initial singularity problem
will most likely require nonperturbative physics, and most certainly will require a better understanding
of time-dependent solutions in string theory.

In this paper we provide a toy model of a nonsingular FRW
cosmology, based on string theory. Our goal is to discuss only one possible way in which string
theory may begin to address the initial singularity problem. Our concrete starting point 
is the low-energy effective action of heterotic string theory including possible 
leading order $\alpha'$ corrections~\footnote{For some other studies of $\alpha'$ corrections
within the context of superstring cosmology see~\cite{Antoniadis:1993jc}-\cite{Tsujikawa:2003pn}.}.  
While the exact structure of these corrections is not known, 
we provide an example in which the resulting cosmology is singularity-free~\footnote{Our 
construction is not designed to remove singularities
that occur when none of the curvature invariants blow up \eg~ the singularity present
in a Taub-NUT space~\cite{hande} or those in~\cite{Horowitz:bv}. We are interested in removing
the initial Big-Bang singularity which is a curvature singularity.}. 
To study the system we use both analytic and numerical techniques.
The metric is asymptotically Minkowski spacetime at low-curvatures and evolves to de Sitter 
space in the high-curvature regime. The significance of the $\alpha'$ corrections
is controlled, in part, by the dilaton field $\P$. These higher-derivative terms become important
in the high-curvature regime and are constructed to allow de Sitter solutions in the 
early universe. In this way our model provides the 
first natural realization of the Limiting Curvature Construction (LCC) in terms of a 
well-motivated, physical theory~\cite{markov}-\cite{easson2003}. Our analysis involves
a dynamical dilaton and a novel, string-inspired form for the higher-derivative terms.
%-----------------------------------------------------------------------  
\section{The Action}
%-----------------------------------------------------------------------  
In $D$-dimensions, the string tree-level effective action for the massless boson
sector (in the string-frame) is:
\bea\label{sf}
\tilde S &=& \frac{\ms^{D-2}}{2} \int d^{D} \! x \sqrt{- \tilde g}\,e^{-2\P} \,\Big\{\, \tilde R 
\,+\, 4 (\nabla \P)^2 \nonumber \\
&+& \z_0 \a' \, \tilde \L2 \,-\frac{2(D-10)}{3\a'}
+ \O(a'^2 \tilde R^4 +\cdots) \, \Big\}
\eea
where $\P$ is the dilaton, $\a'=\ls^2$ and the tilde indicates that we are using the 
string-frame metric $\tilde g_{MN}$, where $M,N = 0,\dots,d$. In the above, $\z_0$
takes on the values $1/4, \, 1/8, \,  0$ for the bosonic, heterotic and type II superstring
theories, respectively. The ``$\cdots$" refer to other higher-derivative terms of order $\a'^2$,
and the Lagrangian $\tilde\L2$ represents the leading-order corrections to the action
and is made up of four-derivative terms 
$\tilde \L2 (\tilde R^2, \, \tricci, \, \triemann, \dots)$ \cite{Gross:1986mw,Metsaev:1987zx}. We choose 
to work within heterotic sting theory.
While the two-point action for the heterotic string includes a gauge vector boson $A_n$ (with
field strength tensor $F_{mn}$) and an anti-symmetric tensor $H_{abc}$:
\bea
\mL_{2pt} &=& \frac{1}{2\k^2}\,R\,-\,\frac{1}{6}\, e^{-2\P}\, H_{abc}H^{abc}\,-\,\half \, (\nabla_c \P)(\nabla^c\P) \nonumber \\
	&\,-\,& \frac{1}{4} \,e^{-\P} \,F_{mn}F^{mn}
\eea
we consider only the most relevant massless modes, the dilaton $\P$ and graviton $g_{mn}$. 
In order to obtain results that are easy 
to compare with General Relativistic theory we work in the Einstein-frame
obtained via a conformal transformation of the metric $\tilde g_{\a\b} = e^{\xi \P} g_{\a\b}$.
Applying this transformation to (\ref{sf}) gives the Einstein-frame action
\bea\label{ef}
S &=& \frac{\ms^{D-2}}{2} \int d^{D} \! x \sqrt{- g}\,\Big\{\, R 
\,-\, \xi (\nabla \P)^2 \nonumber \\
&+& \z_0 \a' e^{-\xi\P}\, \Big[\L2 \,+\,\xi^2\frac{D-4}{D-2} (\nabla \P)^4 \Big] \,-\Lambda e^{\xi\P} \nonumber \\
&+& \O(a'^2 R^4 +\cdots \, ) \, \Big\}
\,,
\eea
where $\xi = 4/(D-2)$ and $\Lambda=2(D-10)/3\a'$ is a contribution to the
cosmological constant that vanishes in ten-dimensions.
In the Einstein-frame the dilaton 
couples directly to the higher-derivative terms $e^{-\xi\P}\L2$. In this way it is possible to control the
``strength" of the $\L2$ terms, in part, via the dilaton. 

Our knowledge of the exact form of $\L2$ is incomplete. Requiring
the action to reproduce string theory S-matrix elements can determine only
some of the coefficients of potential covariant terms in $\L2$. This is because some terms do not
contribute to the S-matrix or provide contributions that overlap in form with those of other
terms \cite{Metsaev:1987zx,Forger:1996vj}. Some sort of off-shell superstring calculation is required to fix the exact coefficients of these terms.
Scattering amplitude
or string worldsheet $\beta$-function calculations predict only the Riemann squared term $\riemann$ in $\L2$.
In \cite{Metsaev:1987zx} Metsaev and Tseytlin fix the contribution
\be\label{scat}
\bar \L2 \,=\, \z_0 \, e^{-2\P}\,\left(\riemann \,+\, \xi^2 \, \frac{D-4}{D-2}\, (\nabla \P)^4 \, \right)
\,,
\ee
where $\bar \L2$ is the term being multiplied by $\a'$ in (\ref{ef}).
However, terms such as $R^2$ and $\ricci$ may be added and subtracted from $\L2$ by performing
field redefinitions of $\mL_{2pt}$ \cite{Gross:1986mw}. Because of this imprecise knowledge of $\L2$, $\L2$ is 
commonly assumed to be the Gauss-Bonnet invariant
\be\label{gaus}
\mL_{GB} = \riemann \,-\, 4\, \ricci \,+\,R^2
\,.
\ee
An advantage of choosing this particular structure for $\L2$ is that the resulting
equations of motion are second-order. The cost
of choosing an invariant of a form other than (\ref{gaus}) is that our equations
of motion will be fourth-order.
Because we are interested in finding a nonsingular
theory we will design something other than the Gauss-Bonnet form for $\L2$~\footnote{Some
examples of nonsingular cosmologies constructed using the Gauss-Bonnet invariant are in
\cite{Kanti:1998jd}.}. A simple way to ensure
that our theory has a nonsingular solution is to choose an $\L2$ that will vanish
for a nonsingular spacetime. We can then look for solutions that approach this nonsingular manifold
in the large curvature regime. As predicted by the string theory
calculation (\ref{scat}), we take the leading term in $\L2$ to be $\riemann$. 
An elementary set of nonsingular spacetimes are the set of maximally symmetric spacetimes 
of constant curvature. The metrics of constant 
curvature are locally characterized by the condition (in $D=4$)
\begin{equation}\label{mki}
R_{\a\b\g\delta} = \frac{1}{12} \, R \,  (\, g_{\a\g}g_{\b\delta} \,-\, g_{\a\delta}g_{\b\g}\,)
\,,
\end{equation}
which is equivalent to 
\begin{equation}\label{i222}
C_{\a\b\g\delta} \,= \, 0 \,=\, R_{\a\b} \,-\, \frac{1}{4} \,R \,g_{\a\b}
\,.
\end{equation}
The space with constant curvature and $R=0$ is Minkowski spacetime.
The space for $R>0$ is de Sitter spacetime, which has topology $\mR^1 \times \mS^3$.
The space with $R<0$ is anti-de Sitter spacetime, which has topology $\mS^1 \times \mR^3$.

Using \eq{mki} it is easy to construct the desired invariant that will vanish 
for spacetimes of constant curvature (in $D=4$) and whose leading term is $\riemann$:
\be\label{i2}
\L2 \,=\,\La2
\,.
\ee
%1
Note that it is possible that there are other solutions satisfying $\L2=0$ (see [47]).
This is a limitation of the construction as presented. If there are other solutions
satisfying this condition then we could try to modify this Lagrangian so that these
solution no longer obey this condition. However, we are not primarily interested in finding
all the solutions to this theory. Our main goal is to show only that 
nonsingular solutions of the type described above exist.
The four-dimensional
action is given by \eq{ef}, and 
may be written
\bea\label{act3}
S &=& \frac{\ms^2}{2} \int d^{4} \! x \sqrt{- g}\,\Big\{\, R\,-\, 2 (\nabla \P)^2 \nonumber \\
&+& D(\P)\, \Big[\La2 \Big] \, - V(\P)\Big\}
\,,
\eea
where $D(\P)= \a' e^{-2 \P}/8$. In the above we have assumed a potential for the
dilaton field $V(\P)$ and we have ignored the contribution to the cosmological
constant term, $\Lambda e^{2\P}$ (\ie{} we do not attempt to solve the 
cosmological constant problem). We will comment further on the dilaton potential below.

Variation of the action with respect to the metric tensor yields:
\begin{widetext}
\begin{eqnarray}
&&4(R_{;\b} \,D(\P)_{;\a}\,  
+ R_{;\a} \,D(\P)_{;\b} + D(\P)_{;\g} \,(12 \, R_{\a\b;}{}{}^\g - 6(R_\a{}^\g{}_{;\b} + R_\b{}^\g{}_{;\a} )-R_;{}^\g\, g_{\a\b})) 
+ 2(2 R\, D(\P)_{;\a\b} \nonumber \\
&+& 2(-5\, R_{;\a\b} \,+\, + 6(R_{\a\b;\g}{}^\g \,+\, R_{\a\b;}{}^\g{}_\g ))\, D(\P) -3(R\,-\,2\P_{;\g} \P_;{}^\g)g_{\a\b} 
+ 6(R_{\a\b} \,+\, 2D( \P)_;{}^{\g\l}R_{\g\b\l\a} \nonumber \\
&+& 2D(\P)_{;\g\l}R_\a{}^\l{}_\b{}^\g)) +  D(\P) (-4 (R R_{\a\b}
+ 12 R_{\g\b} R_\a{}^\g - 12 R_{\g\l} R_\a{}^\g{}_\b{}^\l + 6 R_{\g\l\r\a}R_\b{}^{\r\g\l}) \nonumber \\
&+& g_{\a\b} (R^2 - 4 R_{;\g}{}^\g- 6 R_{\g\l\r\s} R^{\g\l\r\s}))
+ 6 g_{\a\b}   V(\P) \nonumber \\
&\,=\,& 4 R D(\P)_;{}^\g{}_\g \,  g_{\a\b}  \,+\, 4 D(\P)_;{}^\g  \,(6 (R_{\g\a;\b}   \, +\, R_{\g\b;\a}- 2 R_{\a\b;\g}  ) \,+\, R_{;\g}   g_{\a\b}  )
\,,
\label{varyeom}  
\end{eqnarray}     
\end{widetext}
while variation with respect to the field $\P$ gives
\bea\label{eompsi}
&&\sqrt{-g}\Big((\La2) \frac{\d D(\P)}{\d \P}+ 4\nabla^2 \P \Big) \nonumber \\
&=& \sqrt{-g}\frac{\d V(\P)}{\d\P}
\,.
\eea
\section{Cosmological solutions}
We now study cosmological solutions to the theory (\ref{act3}). We assume a time-dependent 
FRW homogeneous and isotropic metric
\be\label{gmn}
ds^2=-n^2(t)dt^2+a^2(t) \left[\frac{dr^2}{1-k{r^2}}+r^{2}d{\theta}^2+r^2 \sin^2{\theta}d\vp^2\right]
\ee
where $a(t)$ is the scale factor of the universe and $n(t)$ is the lapse function. For simplicity we take
the metric to be flat ($k=0$). We also choose a time-dependent, homogeneous dilaton $\P(t)$.
An alternative derivation of the equations of motion (\ref{varyeom}) and (\ref{eompsi}) is achieved
by substituting the metric (\ref{gmn}) into (\ref{act3}) and varying the action with respect
to $a$, $n$, and $\P$. The resulting EOM are:
\begin{widetext}
\bea
e^{2\P} a^4 (V(\P) \,-\, 2 \dot \P^2 ) = 6 \l \dot a^4  &+& 8 \l a \dot a^2  (\,2 \dot a \dot \P \,-\, 3 \ddot a\,)
+ 2 a^2 (\,3 \l \ddot a^2 \,+\, \dot a^2 (\,e^{2 \P} \,-\, 8 \l \dot \P^2 \,+\, 4 \l \ddot \P \,) \nonumber \\
&+& 4 \l \dot a \,\,\ddot{}\! a \dot{} \,)
+ 4 a^3 (\ddot a (e^{2 \P}  + 4 \l \dot \P^2 - 2 \l \ddot \P) + \l ( -4 \dot \P \,\,\ddot{}\! a \dot{} \,+\, \ddot{}\!a \ddot{}\,))  \label{eom1}\\
18 \l \dot a^4  + e^{2 \P} a^4  (V(\P) + 2 \dot \P^2 ) &=&  6 a (2 \l \dot a^2 (2 \dot a \dot \P + \ddot a) +
a (e^{2 \P} \dot a^2  - \l \ddot a^2  + 2 \l \dot a ( -2 \dot \P \ddot a + \,\ddot{}\! a \dot{}\,))) \label{eom2}\\
12 \l (\, \dot a^2  \,-\, a \ddot a\,)^2  &+& e^{2\P}  a^3 \, (\,12 \dot a \dot \P \,+\, a (\,V'(\P) \,+\, 4 \ddot \P\,)\,) \,=\, 0 \label{eom3}
\,,
\eea
\end{widetext}
where $\l=\a'/8$ and the prime in the last equation denotes differentiation with respect to $\P$. In the above
we have set
$n=1$. Note that the dynamics of this system are completely determined by the $G_{00}$ equation (\ref{eom2}) and
the EOM for the scalar field (\ref{eom3}). Equation (\ref{eom1}) serves as a constraint equation. 
To reduce these equations to a second-order system, introduce $H=\dot a/a$. 
Equations (\ref{eom2}) and (\ref{eom3}) reduce to
\bea
\dot H^2 &-& 2H \ddot H -6 H^2 \dot H + 4 H \dot H \dot \P \nonumber \\
&=& -\frac{e^{2\P}}{6 \l} \left(V(\P) + 2\dot \P^2 - 6 \l H^2 \right) \label{alexander}\\
12 \l \dot H^2 &=& - e^{2\P}  \left(\frac{\d V(\P)}{\d\P} + 4 \ddot \P + 12 H \dot \P \right)
\,,
\eea 
%It is instructive to consider the case of constant dilaton $\P=\P_0$. 
%$\x= a^2 \dot a^2$. Equations (\ref{eom2}) and (\ref{eom3}) reduce to
%\bea
%V(\P_0)\,a^4 &+& 6\x - 6 \l e^{2\P_0} \left(\frac{\x\x''}{a^2} - \frac{2\x\x'}{a^3} - \frac{\x'^2}{4a^2}\right) =0 \label{alexander}\\
%\frac{\d V(\P_0)}{\d\P} \, a^4 &+& 12 \l e^{2\P_0}  \left(\frac{\x'^2}{4a^2} - \frac{3}{2}\frac{\x\x'}{a^3} + \frac{3\x^2}{a^4}\right)=0
%\,,
%\eea 
%where prime denotes differentiation with respect to the scale factor $a$.
%
\subsection{\it The late-time universe}
At late times (low curvatures) our action must mimic General Relativity and the Einstein-Hilbert action
$
S_{EH} = (1/2\k^2) \int d^4x \sqrt{-g}\, R
\,.
$
This implies the contribution $D(\P) \L2 - 2 (\nabla \P)^2 - V(\P)$ in equation (\ref{act3}) 
must not produce any physical deviations from $S_{EH}$ in the low-curvature regime. 
Furthermore, the dilaton must approach a $\P \mapsto \P_0$. If the dilaton 
remained time-dependent it would produce observational consequences such as variations in gauge-couplings. 
We also demand $V(\P) \mapsto 0$ to avoid gaining a contribution to the cosmological constant in the late universe.
A de Sitter space solution $a(t) = a_0 \exp{H t}$ corresponds to constant $H=H_0$. In the case under consideration
($V(\P) \mapsto 0$, $\P \mapsto \P_0$) the EOM are satisfied by a de Sitter solution \emph{with the constraint} 
$H=0$ and hence, $a(t)=const$. Therefore, in the (late-time) case of constant dilaton and vanishing potential 
the equations of motion are solved by Minkowski spacetime (the special case of de Sitter with $H=0$)
with flat metric $g_{\a\b} = \eta_{\a\b}=diag(-1, 1, 1, 1)$. 
\subsection{\it The early universe}
In order to remove the initial Big-Bang curvature singularity present in Einstein gravity
at $t=0$, the term $\L2$ must become significant. This will force the metric solution to 
de Sitter spacetime with metric (\ref{gmn}) and $a(t)\propto \exp{H t}$. Discovering a solution which evolves
smoothly from a de Sitter phase at early times to Minkowski at late times will provide an
example of a universe which is everywhere nonsingular. We now produce such a solution and proceed to
study its stability against classical perturbations.
Simple forms for $\P$ and $V(\P)$ that provide the desired 
behavior and obey the EOM are
\be\label{spsi}
\P(t) = \P_0 \tanh{ \left(\frac{t- t_1}{t_0} \right)}
\,,
\ee
and 
\be\label{sv}
V(\P) = V_0 \left((\P+1)^2 - 4 \right)
\,,
\ee
where we set $\P_0 = 1$, $V_0 = -3H_0^2/2$ and $t_0 = 1$. The constant $t_1$ is taken to be large enough
so that $\P(t=0) \approx -1$. Note that the potential (\ref{sv}) is everywhere positive for all allowed values
of the dilaton $\P$ (over the range, $(-1,1)$) and hence, for all time $t \in (-\infty, \infty)$ (see Fig.1). 
\begin{figure}\label{VofP.eps}
\begin{center}
 \epsfxsize=3.0 in \centerline{\epsfbox{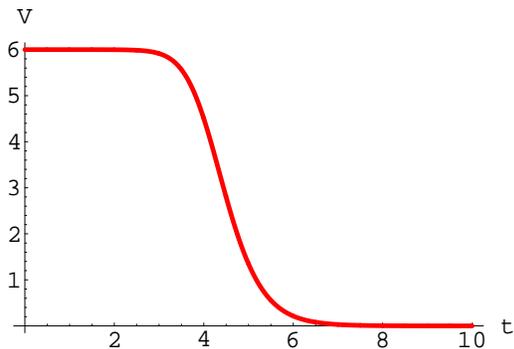}}
  \end{center}
    \caption{The time evolution of the potential $V(\P)$. For this plot (and all subsequent plots) we
    take $t_1 =4$ and $H_0 =1$.}
		    \end{figure}
The choices for $\P$ and $V$ are certainly not unique.
Any $\P$ and $V$ that have the correct asymptotic behavior have the potential to generate nonsingular
solutions of the type under consideration. Inserting the $\P$ ansatz (\ref{spsi})
and the $V$ ansatz (\ref{sv}) into the EOM (\eqto{eom1}{eom3}) we find that as $t \mapsto 0$,
$a(t) \propto \exp{H t}$ (in our specific example $H_0=1$). At late times the metric approaches flat Minkowski space.

\subsection{\it Numerical analysis}
While we have shown that our solution has the desired asymptotic behavior, namely, de Sitter space at
early times and Minkowski space at late times, we must prove that this solution remains nonsingular for
all values of $t$ in between. To do this we conduct a numerical study of the equations of motion. 
Consider the phase space
generated by the equations (\ref{eom2}) and (\ref{eom3}) with $\P$ and $V$ given by (\ref{spsi})  and (\ref{sv}).
The $(\P,H)$-plane is plotted in Fig.2. Note that the range of $\P$, $(-1,1)$ is mapped to the entire
history of the universe $t \in (-\infty, \infty)$. 
		    
Clearly, the curvature is bounded for all time and therefore,
the universe is free of curvature singularities. At early times ($\P \mapsto -1$) the curvature $|H|$ is approximately a positive constant 
corresponding to the de Sitter phase. At late times the dilaton approaches a constant 
($\P \mapsto 1$), the curvature vanishes and the spacetime approaches flat Minkowski space.
%2 OF COURSE THE LATE-TIME UNIVERSE IS NOT MINKOWSKI SPACETIME...
Of course, recent observational evidence from supernovae data~\cite{Riess:1998cb} and the WMAP satellite~\cite{Spergel:2003cb}
indicate that the universe is not only expanding but undergoing an accelerated expansion. It is easy
to modify the Minkowski condition in this model in order to accommodate various late time behaviors
(for example, by adding an appropriate matter Lagrangian to the system). To achieve an accelerating
universe at late times one simply could tune the potential to not go exactly to zero at late times
like in quintessence models.

It is important to discuss the stability of this solution. We now show that this solution
is not (in general) an attractor at early times but is at late times. Therefore, nonsingular solutions
of the type described here are not generic in the model under consideration. It is possible, however,
that other nonsingular solutions exist in this theory that are attractors. Furthermore,
as pointed out by Starobinsky~\cite{starob}, the evolution of the Universe need
not follow a generic solution, it may well be described just
by this unique one, at least initially. The late time behavior of 
Minkowski spacetime with frozen dilaton is classically stable against small perturbations.  
Let us begin with the stability analysis of the early-time
solution. For our considerations it is sufficient to consider stability against homogeneous fluctuations. 
An analysis of inhomogeneous fluctuations is considerably more involved.
\begin{figure}\label{HP.eps}
\begin{center}
 \epsfxsize=3.0 in \centerline{\epsfbox{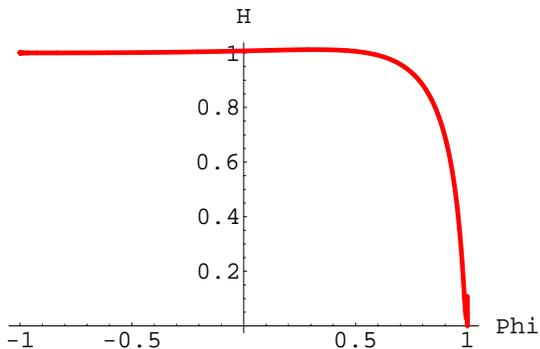}}
  \end{center}
    \caption{The $(\P,H)$-phase portrait generated from the equations of motion. Note
    that the curvature remains finite for the entire history of the universe.}
		    \end{figure}
Let
\be
a(t) = a_0 e^{Ht} \left(1+\delta(t)\right), \qquad |\delta(t)|<<1
\,.
\ee
\be\label{pper}
\P(t) = \P_0 \tanh{ \left(\frac{t- t_1}{t_0} \right)} \left(1+\epsilon(t)\right), \qquad |\epsilon(t)|<<1
\,.
\ee
Recall, that when $t \mapsto 0$, $\P \mapsto -1$. To first approximation in $\delta$ and $\epsilon$
the $G_{00}$ equation and the equation of motion for $\P$ become
\bea
\ddot \delta &+& 3 H \dot \delta + \frac{\delta}{\l e^2} = 0 \, \\
4 \ddot \epsilon &+& 12H\dot\epsilon = 3H^2 \epsilon
\,.
\eea
The solutions to these equations are
\be
\delta = \frac{d_1 + d_2 \, e^{\frac{\sqrt{9 \l e^2 H^2 -4} \,t}{\l^{1/2} e}}}{\exp{(((3He + \frac{\sqrt{9 \l e^2 H^2 -4}}{\l^{1/2}}))t/2e)}}
\ee
\be
\epsilon = \frac{c_1 + c_2 \, e^{2\sqrt{3} \,Ht}}{e^{((3 + 2 \sqrt{3}) \, Ht)/2}}
\,.
\ee
Here $d_1$, $d_2$, $c_1$ and $c_2$ are constants. The conditions that $\epsilon$ and $\delta$ be
small ($<<1$) at the beginning of our analysis when $t \simeq 0$ implies that $|c_1 + c_2|<<1$ and $|d_1 + d_2|<<1$. 
While $\delta$ remains small for all time (for small $d_1$ and $d_2$) $\epsilon$, in general, grows greater than unity
and our solution becomes unstable~\footnote{For $c_2 =0$ the solution is stable.}.  Hence, at early times the solution
is stable against perturbations of the metric but not against perturbations of the dilaton $\P$.
This demonstrates that nonsingular solutions of the type given above are not generic.

To see if the late-time Minkowski space solution is stable we consider perturbations around the Minkowski
solution 
\be
a(t) = a_0 \left(1+\delta(t)\right), \qquad |\delta(t)|<<1
\,,
\ee
and perturbations of the dilaton of the form (\ref{pper}).
Recall that at late times (as $t  \mapsto \infty$), $\P \mapsto 1$. To first approximation in $\delta$ and $\epsilon$
the $G_{00}$ and the equation of motion for $\P$ give $\delta=\epsilon=0$. Therefore, our late-time
Minkowski space solution is an attractor.

\section{Conclusions}
%------------------------------------------------------------------------------------------
%------------------------------------------------------------------------------------------
\label{conc}  
%------------------------------------------------------------------------------------------
In this letter we studied cosmological solutions to heterotic string theory including
a possible form for leading order $\a'$ corrections to the low-energy effective action. A nonsingular solution
is given in which the universe evolves from an early-time de Sitter phase to a late-time Minkowski spacetime with
constant dilaton. One limitation of this model is that at high-energies and large curvatures
$E^2\a' \simeq \O(1)$ higher-order
curvature corrections (\eg{}  $\a'^2$ corrections) become important. Including these terms or quantum loop corrections 
in $\gs = e^{\left<\P \right>}$
could, presumably, re-introduce the singularity. Of course, if the string scale is TeV then such corrections 
will become important sooner then if the string scale is near the Planck scale.
It is likely that a full nonperturbative analysis is required in order
to fully understand the initial singularity problem. Our analysis is meant to show a possible way
in which string theory may address this issue. We have shown that including higher-derivative terms in 
the action can (under certain circumstances) resolve the initial curvature singularity.

Finally, it would be interesting to see if this construction can regulate the bounce that
occurs in the Pre-Big-Bang (PBB) model of string cosmology~\cite{Veneziano:1991ek, Gasperini:1992em}. 
The analysis would involve ideas from this paper and from \cite{Brandenberger:1998zs, Easson:1999xw}.
%3.COMMENTS ABOUT PBB...
In the PBB model the universe starts out in the perturbative string Minkowski vacuum. The universe
then goes through a high-curvature, collapsing regime during which the $\a'$ corrections become important.
Hopefully, after a successful ``graceful exit", the universe enters an expanding low-curvature radiation
(or quintessence) dominated phase. In this paper we have studied only the post-collapse branch of the
PBB model. The collapsing branch would be the time reverse of this. We leave a more detailed analysis 
of the PBB scenario in this context to a future paper.
\begin{acknowledgments}  
%----------------------------------------------------------------------- 
It is a pleasure to thank R.~Brandenberger, C.~Burgess, H.~Firouzjahi, P.~Martineau
and A.~Mazumdar for helpful discussions.  
%----------------------------------------------------------------------- 
\end{acknowledgments}

\end{document}